\documentclass[a4paper,10pt,eqno]{article}
\usepackage{theorem}
\usepackage{latexsym,amssymb,amsfonts,amsmath}
\newtheorem{thm}{Theorem}[section]
\newtheorem{lemm}[thm]{Lemma}

\newtheorem{pro}[thm]{Proposition}

\newtheorem{defi}{Definition}[section]

\theoremheaderfont{\scshape}

\begin{document}
\title{{\Large {\bf LOCALIZATION DOES NOT OCCUR FOR THE FOURIER WALK ON THE MULTI-DIMENSIONAL LATTICE}}}
\author{
{\small Akihiro Narimatsu\footnote{narimatsu-akihiro-pd@ynu.jp(e-mail of the author)}}\\
{\scriptsize  Department of Applied Mathematics, Graduate School of Engineering Science, Yokohama National University}\\
{\scriptsize \footnotesize\it 79-5 Tokiwadai, Hodogaya, Yokohama, 240-8501, Japan}\\
}
\date{}
\maketitle
\par\noindent
\begin{small}
\par\noindent
{\bf Abstract}. The existence of localization for the Grover walk on the multi-dimensional lattice is known. This paper gives some conditions for the existence of localization for the space-homogeneous quantum walks. We also prove that localization does not occur for the Fourier walk on the multi-dimensional lattice.
\end{small}
\section{Introduction}

The quantum walks (QWs) were introduced by Aharonov et al. \cite{Aharonov1} as the quantum version of the usual random walks. QWs have been intensively studied from various fields, such as quantum algorithm \cite{Portugal}, the topological insulator \cite{Kitagawa}, and radioactive waste reduction \cite{Matsuoka}.

The properties of QWs in one dimension, especially the Hadamard walk, are well studied \cite{Konno1}. However, properties of QWs in higher dimensions have not been clarified except for the Grover walks \cite{komatsukonno, konnotakahashi, stefanak, watabeetal}.

Although the Fourier walk includes the Hadamard walk as the one-dimensional version of it, results of the Fourier walk on the multi-dimensional case are limited. Komatsu and Tate \cite{Komatsu and Tate.} showed the non-existence of localization for the 2-dimensional lattice, and Asano et al. \cite{asanoetal} gave another proof of it. These proofs have in common that they need the characteristic polynomial of evolution operator in order to show the non-existence of the constant eigenvalue of the evolution operator.

In this paper, we get a necessary and sufficient condition and a necessary condition for the existence of localization for space homogeneous QWs on higher-dimensional lattice. And our main result is that localization does not occur for the Fourier walk on the $d$-dimensional lattice $(d=1,2,3,\dots)$. Compared with the above-mentioned previous result, we obtain general results for the Fourier walk, and also get some claims for general space-homogeneous QWs. Moreover, our proof is simpler and more applicable because we don't need the characteristic polynomial of the evolution operator.

The rest of the paper is organized as follows. Section 2 is devoted to the definition of the discrete time QWs. Section 3 presents our results. We show that localization does not occur for the Fourier walk on the multi-dimensional lattice. Section 4 summarizes our paper.
\section{Definitions}
In this section, we present the definitions of our model.
\subsection{QWs on $\mathbb{Z}^d$}
In this subsection, we define QWs on $\mathbb{Z}^d(d=1,2,\dots)$. We introduce the Hilbert space as follows:
\begin{align*}
{\mathcal H}=\ell^2(\mathbb{Z}^d\otimes\mathbb{C}^{2d})=\{\Psi:\mathbb{Z}^d\to\mathbb{C}^{2d}\ |\ \sum_{x\in\mathbb{Z}^d}\|\Psi(x)\|^2<\infty\}.
\end{align*}
Next let $I_{d}$ be the identity map on $\mathbb Z^d$, coin operator $C$ be a unitary matrix with size $2d$ and shift operator $S$ be
\begin{align}
S=\sum_{x\in \mathbb Z^d}\sum_{j=0}^{d-1}\bigl(|x-e_j\rangle\langle x|\otimes |2j\rangle\langle 2j|+|x+e_j\rangle\langle x|\otimes |2j+1\rangle\langle 2j+1|\bigr), \label{shiftop}
\end{align}
where $e_j$ is a normalized basis of $j$-th axis. Then the evolution operator is given by
\begin{align}
U=S(I_d\otimes C).\label{evoop}
\end{align}
We put $\Psi_n$ as the state of QWs at time $n\in\{1,2,3,\dots\}$. Thus $\Psi_n$ can be written as follows:
\begin{align}
\Psi_n=U^n\Psi_0,\label{timeevo}
\end{align}
where $\Psi_0\in\mathcal H$ is the initial state of QWs. We should remark that by Eqs.(\ref{shiftop},\ref{evoop},\ref{timeevo}), we have another description for time evolution of QWs: 
\begin{align}
\Psi_{n+1}(x)=\sum_{j=0}^{d-1}|2j\rangle\langle 2j|C\Psi_n(x+e_j)+|2j+1\rangle\langle 2j+1|C\Psi_n(x-e_j).\label{timeevo2}
\end{align}

In this study, we mainly consider the Fourier walk on $\mathbb Z^d$ with the coin operator:
\begin{align}
C_d=\frac{1}{\sqrt{2d}}\begin{bmatrix}
1&1&1&\cdots&1\\
1&\omega_{2d}&\omega_{2d}^2&\cdots&\omega_{2d}^{2d-1}\\
1&\omega_{2d}^2&\omega_{2d}^{2\cdot2}&\cdots&\omega_{2d}^{2\cdot(2d-1)}\\
\vdots&\vdots&\vdots&\ddots&\vdots\\
1&\omega_{2d}^{2d-1}&\omega_{2d}^{(2d-1)\cdot2}&\cdots&\omega_{2d}^{(2d-1)\cdot(2d-1)}
\end{bmatrix},\label{fouriercoin}
\end{align}
where $\omega_j=\exp(2\pi i/j)$.

\subsection{Localization}

This subsection deals with definitions of localization and the Fourier analysis.
\vspace{1\baselineskip}
\begin{defi}
Localization of QWs occurs if and only if there exists $\Psi_0\in\mathcal H$ such that $\displaystyle{\overline{\lim_{n\to\infty}}\|\Psi_n(x)\|>0}$ is satisfied at some $x\in\mathbb Z^d$.
\end{defi}

Next we introduce the Fourier analysis. Let $\Hat\Psi_n(k)$ be
\begin{align*}
\Hat\Psi_n(k)=({\mathcal F}\Psi_n)(k)=\sum_{x\in\mathbb Z^d}e^{-i\langle k,x\rangle}\Psi_n(x),
\end{align*}
where $k=(k_1,k_2,\dots,k_d)\in[0,2\pi)^d$. We should note that
\begin{align*}
\Psi_n(x)=({\mathcal F}^{-1}\Hat\Psi_n)(x)=\int_{[0,2\pi)^d}e^{i\langle k,x\rangle}\Psi_n(k)\frac{{\rm d}k}{(2\pi)^d}.
\end{align*}
The time evolution of QW on $k$-space is written as
\begin{align*}
\Hat\Psi_{n+1}(k)=\Hat U(k)\Hat\Psi(k),
\end{align*}
where
\begin{align*}
\Hat U(k)=\begin{bmatrix}
e^{ik_1}&0&\cdots&0&0\\
0&e^{-ik_1}&\cdots&0&0\\
\vdots&\vdots&\ddots&\vdots&\vdots\\
0&0&\cdots&e^{ik_d}&0\\
0&0&\cdots&0&e^{-ik_d}
\end{bmatrix}C.
\end{align*}
Then the following proposition is obtained in \cite{Tate}.
\vspace{1\baselineskip}
\begin{pro}\label{tatens}
Localization of QWs occurs if and only if $\Hat U(k)$ has a constant eigenvalue. 
\end{pro}

\section{Results}
Our purpose in this section is to prove the following main result:
\vspace{1\baselineskip}
\begin{thm}\label{main}
Localization does not occur for the Fourier walk on $\mathbb Z^d\ (d=1,2,3,\dots)$. 
\end{thm}
\vspace{1\baselineskip}
In subsection \ref{sec:condiofloc}, we get some lemmas.
\subsection{Conditions of localization}\label{sec:condiofloc}
In this subsection, we consider some conditions of localization. To state Lemma \ref{nscondi}, we introduce the following sets. One is the set of solutions of the eigenvalue problem:
\begin{align*}
{\mathcal W}^{(\lambda)}=\{\Psi^{(\lambda)}\in{\mathcal H}\setminus 0\ |\ U\Psi^{(\lambda)}=\lambda\Psi^{(\lambda)}\},
\end{align*}
for $\lambda\in\mathbb C$ satisfying $|\lambda|=1$. The other is the set of states with finite support:
\begin{align*}
{\mathcal S}_f=\bigl\{\Psi\in{\mathcal H}\ |\ \#\{x\in\mathbb Z^d\ |\ \Psi(x)\neq0\}<\infty\bigr\}.
\end{align*}
\begin{lemm}\label{nscondi}
A necessary and sufficient condition on the existence of localization for space-homogeneous QWs on $\mathbb Z^d$ is the following:
\begin{align*}
{\rm There\ exists}\ \lambda\in\mathbb C\ {\rm such\ that}\ |\lambda|=1\ {\rm and}\ {\mathcal W}^{(\lambda)}\cap{\mathcal S}_f\neq\emptyset.
\end{align*}
\end{lemm}
{\it Proof}.\ The sufficiency is clear. Therefore we consider the necessity. By Proposition\ \ref{tatens}, $\Hat U(k)$ has an eigenvalue $\lambda\in\mathbb C$, which does not depend on $k$. Then it suffices to show that ${\mathcal W}^{(\lambda)}\cap {\mathcal S}_f\neq\emptyset$. Let $v^{(\lambda)}(k)=\!^T\begin{bmatrix}v_0^{(\lambda)}(k)&v_1^{(\lambda)}(k)&\dots&v_{2d-1}^{(\lambda)}(k)\end{bmatrix}\in\mathbb C^{2d}$ be the eigenvector of $\Hat U(k)$, which corresponds to $\lambda$. We should note that
\begin{align*}
\Hat U(k)v^{(\lambda)}(k) =\lambda v^{(\lambda)}(k).
\end{align*}
Then, there exists $\ell\in\{0,1,\dots,2d-1\}$ such that each of $j\in\{0,1,\dots,2d-1\}$ satisfies
\begin{align*}
v_j^{(\lambda)}(k)=\frac{f_j^{(\lambda)}(e^{ik_1},e^{ik_2},\dots,e^{ik_d})}{g_j^{(\lambda)}(e^{ik_1},e^{ik_2},\dots,e^{ik_d})}v_\ell^{(\lambda)}(k),
\end{align*}
where $f_j^{(\lambda)}(x_1,x_2,\dots,x_d)$, and $g_j^{(\lambda)}(x_1,x_2,\dots,x_d)\neq0$ are multivariable polynomials of $(x_1,x_2,\dots,x_d)$. Let $v^{(\lambda)}_\ell(k)={\displaystyle \prod_{m=0}^{2d-1}g_m^{(\lambda)}(e^{ik_1},e^{ik_2},\dots,e^{ik_d})}$, we obtain
\begin{align*}
v^{(\lambda)}_j(k)=\frac{f_j^{(\lambda)}(e^{ik_1},e^{ik_2},\dots,e^{ik_d})\prod_{m=0}^{2d-1}g_m^{(\lambda)}(e^{ik_1},e^{ik_2},\dots,e^{ik_d})}{g_j^{(\lambda)}(e^{ik_1},e^{ik_2},\dots,e^{ik_d})}.
\end{align*}
Since $v_j^{(\lambda)}(k)$ is a multivariable polynomial, we have $\Psi^{(\lambda)}=({\mathcal F}^{-1}v^{(\lambda)})\in{\mathcal W}^\lambda\cap{\mathcal S}_f$. Then let $\Psi_0=\Psi^{(\lambda)}$, the necessity has shown.\hfill $\Box$
\vspace{1\baselineskip}

We consider the Grover walk on the $2$-dimensional lattice because it is an example of existence of localization. The coin matrix is given by
\begin{align*}
C=\frac{1}{2}\begin{bmatrix}
-1&1&1&1\\
1&-1&1&1\\
1&1&-1&1\\
1&1&1&-1
\end{bmatrix}.
\end{align*}
According to \cite{stefanak}, an eigenvalue of $\Hat{U}(k)$ is $\lambda=1$, and the corresponding eigenvector is
\begin{align*}
v^{(1)}(k)=\frac{1}{2\sqrt2}^T\!\begin{bmatrix}
1+e^{-ik_2}&e^{-ik_1}+e^{-ik_1-ik_2}&1+e^{-ik_1}&e^{-ik_2}+e^{-ik_1-ik_2}
\end{bmatrix}.
\end{align*}
And we get $\Psi^{(1)}\in{\mathcal W}^{(1)}\cap{\mathcal S}_f$:
\begin{align*}
\Psi^{(1)}=({\mathcal F}^{-1}v^{(1)})=\frac{1}{2\sqrt2}&\Bigl(\ ^T\!\begin{bmatrix}1&0&1&0 \end{bmatrix}\otimes|0,0\rangle+\ ^T\!\begin{bmatrix}0&1&1&0 \end{bmatrix}\otimes|1,0\rangle\\&+\ ^T\!\begin{bmatrix}1&0&0&1 \end{bmatrix}\otimes|0,1\rangle+\ ^T\!\begin{bmatrix}0&1&0&1 \end{bmatrix}\otimes|1,1\rangle \Bigr).
\end{align*}
Konno and Takahashi \cite{konnotakahashi} reported some results on ${\mathcal W}^{(1)}\cap{\mathcal S}_f$ of the Grover walk in higher dimensions.

Using Lemma \ref{nscondi}, we get next lemma.
\begin{lemm}\label{ncondi}
A necessary condition on the existence of localization for space-homogeneous QWs on $\mathbb Z^d$ is the following:
\begin{align*}
{\rm For\ any}\ \ell=(\ell_0,\ell_1,\dots,\ell_{d-1})\in\{0,1\}^d,\ {\rm we\ have\ rank}(C^{(\ell)})<d,
\end{align*}
where $C=\begin{bmatrix}c_{j,k} \end{bmatrix}_{j,k=0,1,\dots,2d-1},\ C^{(\ell)}=\begin{bmatrix}c^{(\ell)}_{j,k} \end{bmatrix}_{j,k=0,1,\dots,d-1}$ with $c^{(\ell)}_{j,k}=c_{2j+\ell_j,2k+\ell_k}$.
\end{lemm}
Before the proof, we consider the following example. For $d=3, \ell=(0,1,0)$, we get
\begin{align*}
C^{(\ell)}=\begin{bmatrix}c_{0,0}&c_{0,3}&c_{0,4}\\c_{3,0}&c_{3,3}&c_{3,4}\\c_{4,0}&c_{4,3}&c_{4,4} \end{bmatrix}.
\end{align*}
{\it Proof of Lemma\ \ref{ncondi}.\ }We consider $\mathbb Z^2$ case, because it is essential. By Lemma \ref{nscondi}, existence of localization ensures that we can choose $\Psi_0\in{\mathcal W}^{(\lambda)}\cap {\mathcal S}_f$, where $\lambda\in\mathbb C$ is an eigenvalue of the evolution operator $U$ with $|\lambda|=1$. Then there exists $(x_0^{(\ell)},x_1^{(\ell)})\in\mathbb Z^2\ \bigl(\ell=(\ell_0,\ell_1)\in\{0,1\}^2\bigr)$ satisfying $\Psi_n(x_0^{(\ell)},x_1^{(\ell)})\neq\bf 0$ and
\begin{align}
\Psi_n(x_0^{(\ell)}-(-1)^{\ell_0},x_1^{(\ell)})=\Psi_n(x_0^{(\ell)},x_1^{(\ell)}-(-1)^{\ell_1})={\bf 0}\ (n=0,1,2,\dots),\label{psi0soba}
\end{align}
since $\Psi_0\in{\mathcal S}_f$. We put $\Psi_n(x_0^{(\ell)},x_1^{(\ell)})=^T\!\begin{bmatrix}\alpha_n^{(\ell)}&\beta_n^{(\ell)}&\gamma_n^{(\ell)}&\delta_n^{(\ell)} \end{bmatrix}$. In order to clarify our argument, we consider $\ell=(0,1)$ case. By Eq.\eqref{psi0soba}, there exists $(x_0^{(0,1)},x_1^{(0,1)})\in \mathbb Z^2$ such that $\Psi_0$ satisfies $\Psi_0(x_0^{(0,1)},x_1^{(0,1)})\neq{\bf 0}$ and 
\begin{align}
\Psi_0(x_0^{(0,1)}-(-1)^0,x_1^{(0,1)})=\Psi_0(x_0^{(0,1)},x_1^{(0,1)}-(-1)^1)={\bf 0}. \label{2dmawari0}
\end{align}

Using Eq.(\ref{timeevo2}), we have 
\begin{align*}
\Psi_1(x_0^{(0,1)},x_1^{(0,1)})&=^T\!\begin{bmatrix}\alpha_1^{(0,1)}&\beta_1^{(0,1)}&\gamma_1^{(0,1)}&\delta_1^{(0,1)} \end{bmatrix}\\
&=\bigl(U\Psi_0\bigr)(x_0^{(0,1)},x_1^{(0,1)})\\
&=|0\rangle\langle0|C\Psi_0(x_0^{(0,1)}+1,x_1^{(0,1)})+|1\rangle\langle1|C\Psi_0(x_0^{(0,1)}-1,x_1^{(0,1)})\\&+|2\rangle\langle2|C\Psi_0(x_0^{(0,1)},x_1^{(0,1)}+1)+|3\rangle\langle3|C\Psi_0(x_0^{(0,1)},x_1^{(0,1)}-1).
\end{align*}
Thus we get
\begin{align}
\beta_1^{(0,1)}&=\langle1|C\Psi_0(x_0^{(0,1)}-1,x_1^{(0,1)}),\notag\\
\gamma_1^{(0,1)}&=\langle2|C\Psi_0(x_0^{(0,1)},x_1^{(0,1)}+1).\label{beta1gamma1}
\end{align}
Substitute Eq.\eqref{2dmawari0} into Eq.(\ref{beta1gamma1}), we obtain $\beta_1^{(0,1)}=\gamma_1^{(0,1)}=0$. Then for any $n\in\{0,1,2,\dots\}$, $\beta_n^{(0,1)}=\gamma_n^{(0,1)}=0$, since $\Psi_0\in{\mathcal W}^{(\lambda)}$. Note that Eq.\eqref{psi0soba} gives $\Psi_n(x_0^{(\ell)}-1,x_1^{(\ell)})=\Psi_n(x_0^{(\ell)},x_1^{(\ell)}+1)={\bf 0}$. From Eq.\eqref{timeevo2}, the following equation:
\begin{align*}
|0\rangle\langle 0|C\Psi_0(x_0^{(0,1)},x_1^{(0,1)})=|3\rangle\langle 3|C\Psi_0(x_0^{(0,1)},x_1^{(0,1)})={\bf 0}
\end{align*} 
is required. Computing this, we have
\begin{align}
\bigl(|0\rangle\langle 0|+|3\rangle\langle 3|\bigr)C\begin{bmatrix}\alpha_0^{(0,1)}\\0\\0\\\delta_0^{(0,1)} \end{bmatrix}=\begin{bmatrix}c_{0,0}&c_{0,1}&c_{0,2}&c_{0,3}\\0&0&0&0\\0&0&0&0\\c_{3,0}&c_{3,1}&c_{3,2}&c_{3,3} \end{bmatrix}\begin{bmatrix}\alpha_0^{(0,1)}\\0\\0\\\delta_0^{(0,1)} \end{bmatrix}={\bf 0}.\label{2dnecekai}
\end{align}
Since Eq.\eqref{2dnecekai} has a trivial solution $\alpha_0^{(0,1)}=\delta_0^{(0,1)}=0$ and $\Psi_0(x_0^{(0,1)},x_1^{(0,1)})\neq{\bf 0}$, we obtain a necessary condition on the existence of localization as follows:
\begin{align*}
{\rm rank}\Bigl(\begin{bmatrix}c_{0,0}&c_{0,3}\\c_{3,0}&c_{3,3} \end{bmatrix}\Bigr)={\rm rank}(C^{(0,1)})<2.
\end{align*}
In a similar fashion, for all $\ell\in\{0,1\}^2$, we have rank$(C^{(\ell)})<2$. We can apply this argument to other dimensions. Then, if localization occurs for a QW on the $d$-dimensional lattice $(d=1,2,3,\dots)$, then we have rank$(C^{(\ell)})<d$ for all $\ell\in\{0,1\}^d$.\hfill $\Box$

\subsection{Proof of Theorem \ref{main}}

By Lemma \ref{ncondi}, it suffices to say that there exists $\ell\in\{0,1\}^d$ such that rank$(C^{(\ell)})=d$. There are two cases for the Fourier walk on the $d$-dimensional lattice, {\bf (i)} $d$ is odd and {\bf (ii)} $d$ is even. Then we consider two cases respectively.\\
{\bf (i)}\ $d$ is odd.

Let $\ell_{odd}\in\{0,1\}^d$ be $\{0,0,\dots,0\}$. We should remark that the coin matrix is given by Eq.\eqref{fouriercoin}:
\begin{align*}
C_d=\frac{1}{\sqrt{2d}}\begin{bmatrix}
1&1&1&\cdots&1\\
1&\omega_{2d}&\omega_{2d}^2&\cdots&\omega_{2d}^{2d-1}\\
1&\omega_{2d}^2&\omega_{2d}^{2\cdot2}&\cdots&\omega_{2d}^{2\cdot(2d-1)}\\
\vdots&\vdots&\vdots&\ddots&\vdots\\
1&\omega_{2d}^{2d-1}&\omega_{2d}^{(2d-1)\cdot2}&\cdots&\omega_{2d}^{(2d-1)\cdot(2d-1)}
\end{bmatrix}.
\end{align*}
Thus we have
\footnotesize
\begin{align*}
C^{(\ell_{odd})}=\begin{bmatrix}
1&1&\cdots&1&1&1&\cdots&1\\
1&\omega_{2d}^{2\cdot2}&\cdots&\omega_{2d}^{2\cdot(d-1)}&\omega_{2d}^{2\cdot(d+1)}&\omega_{2d}^{2\cdot(d+3)}&\cdots&\omega_{2d}^{2\cdot (2d-2)}\\
\vdots&\vdots&\ddots&\vdots&\vdots&\vdots&\ddots&\vdots\\
1&\omega_{2d}^{(d-1)\cdot2}&\cdots&\omega_{2d}^{(d-1)\cdot(d-1)}&\omega_{2d}^{(d-1)\cdot(d+1)}&\omega_{2d}^{(d-1)\cdot(d+3)}&\cdots&\omega_{2d}^{(d-1)\cdot (2d-2)}\\
1&\omega_{2d}^{(d+1)\cdot2}&\cdots&\omega_{2d}^{(d+1)\cdot(d-1)}&\omega_{2d}^{(d+1)\cdot(d+1)}&\omega_{2d}^{(d+1)\cdot(d+3)}&\cdots&\omega_{2d}^{(d+1)\cdot (2d-2)}\\
1&\omega_{2d}^{(d+3)\cdot2}&\cdots&\omega_{2d}^{(d+3)\cdot(d-1)}&\omega_{2d}^{(d+3)\cdot(d+1)}&\omega_{2d}^{(d+3)\cdot(d+3)}&\cdots&\omega_{2d}^{(d+3)\cdot (2d-2)}\\
\vdots&\vdots&\ddots&\vdots&\vdots&\vdots&\ddots&\vdots\\
1&\omega_{2d}^{(2d-2)\cdot2}&\cdots&\omega_{2d}^{(2d-2)\cdot(d-1)}&\omega_{2d}^{(2d-2)\cdot(d+1)}&\omega_{2d}^{(2d-2)\cdot(d+3)}&\cdots&\omega_{2d}^{(2d-2)\cdot (2d-2)}
\end{bmatrix}.
\end{align*}
\normalsize
Noting that $\omega_{2d}^{2d}=1$, we get
\begin{align*}
C^{(\ell_{odd})}=\begin{bmatrix}
1&1&\cdots&1&1&1&\cdots&1\\
1&\omega_{2d}^{2\cdot2}&\cdots&\omega_{2d}^{2\cdot(d-1)}&\omega_{2d}^{2\cdot(d+1)}&\omega_{2d}^{2\cdot(d+3)}&\cdots&\omega_{2d}^{2\cdot (2d-2)}\\
\vdots&\vdots&\ddots&\vdots&\vdots&\vdots&\ddots&\vdots\\
1&\omega_{2d}^{(d-1)\cdot2}&\cdots&\omega_{2d}^{(d-1)\cdot(d-1)}&\omega_{2d}^{(d-1)\cdot(d+1)}&\omega_{2d}^{(d-1)\cdot(d+3)}&\cdots&\omega_{2d}^{(d-1)\cdot (2d-2)}\\
1&\omega_{2d}^{1\cdot2}&\cdots&\omega_{2d}^{1\cdot(d-1)}&\omega_{2d}^{1\cdot(d+1)}&\omega_{2d}^{1\cdot(d+3)}&\cdots&\omega_{2d}^{1\cdot (2d-2)}\\
1&\omega_{2d}^{3\cdot2}&\cdots&\omega_{2d}^{3\cdot(d-1)}&\omega_{2d}^{3\cdot(d+1)}&\omega_{2d}^{3\cdot(d+3)}&\cdots&\omega_{2d}^{3\cdot (2d-2)}\\
\vdots&\vdots&\ddots&\vdots&\vdots&\vdots&\ddots&\vdots\\
1&\omega_{2d}^{(d-2)\cdot2}&\cdots&\omega_{2d}^{(d-2)\cdot(d-1)}&\omega_{2d}^{(d-2)\cdot(d+1)}&\omega_{2d}^{(d-2)\cdot(d+3)}&\cdots&\omega_{2d}^{(d-2)\cdot (2d-2)}
\end{bmatrix}.
\end{align*}
Then $C^{(\ell_{odd})}$ can be transformed by fundamental matrices as follows:
\begin{align*}
\overline C^{(\ell_{odd})}=\begin{bmatrix}
1&1&1&\cdots&1\\
1&\omega_{2d}^{1\cdot2}&\omega_{2d}^{1\cdot4}&\cdots&\omega_{2d}^{1\cdot(2d-2)}\\
1&\omega_{2d}^{2\cdot2}&\omega_{2d}^{2\cdot4}&\cdots&\omega_{2d}^{2\cdot(2d-2)}\\
\vdots&\vdots&\vdots&\ddots&\vdots\\
1&\omega_{2d}^{(d-1)\cdot2}&\omega_{2d}^{(d-1)\cdot4}&\cdots&\omega_{2d}^{(d-1)\cdot(2d-2)}
\end{bmatrix}.
\end{align*}
Since $\overline C^{(\ell)}$ is the Vandermonde matrix with size $d$, we obtain rank$(\overline C^{(\ell_{odd})})=d$, so we conclude rank$(C^{(\ell_{odd})})=d$. \\
{\bf (ii)}\ $d$ is even.

Let $\ell_{even}=\{\ell_0,\ell_1,\dots,\ell_{d/2-1},\ell_{d/2},\ell_{d/2+1},\dots,\ell_{d-1}\}\in\{0,1\}^d$ be $\{0,0,\dots,0,1,1,\dots,1\}$. Then we obtain
\footnotesize
\begin{align*}
C^{(\ell_{even})}&=\begin{bmatrix}
1&1&\cdots&1&1&1&\cdots&1\\
1&\omega_{2d}^{2\cdot2}&\cdots&\omega_{2d}^{2\cdot(d-2)}&\omega_{2d}^{2\cdot (d+1)}&\omega_{2d}^{2\cdot(d+3)}&\cdots&\omega_{2d}^{2\cdot (2d-1)}\\
\vdots&\vdots&\ddots&\vdots&\vdots&\vdots&\ddots&\vdots\\
1&\omega_{2d}^{(d-2)\cdot2}&\cdots&\omega_{2d}^{(d-2)\cdot(d-2)}&\omega_{2d}^{(d-2)\cdot(d+1)}&\omega_{2d}^{(d-2)\cdot(d+3)}&\cdots&\omega_{2d}^{(d-2)\cdot (2d-1)}\\
1&\omega_{2d}^{(d+1)\cdot2}&\cdots&\omega_{2d}^{(d+1)\cdot(d-2)}&\omega_{2d}^{(d+1)\cdot(d+1)}&\omega_{2d}^{(d+1)\cdot(d+3)}&\cdots&\omega_{2d}^{(d+1)\cdot (2d-1)}\\
1&\omega_{2d}^{(d+3)\cdot2}&\cdots&\omega_{2d}^{(d+3)\cdot(d-2)}&\omega_{2d}^{(d+3)\cdot(d+1)}&\omega_{2d}^{(d+3)\cdot(d+3)}&\cdots&\omega_{2d}^{(d+3)\cdot (2d-1)}\\
\vdots&\vdots&\ddots&\vdots&\vdots&\vdots&\ddots&\vdots\\
1&\omega_{2d}^{(2d-1)\cdot2}&\cdots&\omega_{2d}^{(2d-1)\cdot(d-2)}&\omega_{2d}^{(2d-1)\cdot(d+1)}&\omega_{2d}^{(2d-1)\cdot(d+3)}&\cdots&\omega_{2d}^{(2d-1)\cdot (2d-1)}
\end{bmatrix}\\
&=\begin{bmatrix}
1&1&\cdots&1&1&1&\cdots&1\\
1&\omega_{2d}^{2\cdot2}&\cdots&\omega_{2d}^{2\cdot(d-2)}&\omega_{2d}^{2\cdot 1}&\omega_{2d}^{2\cdot 3}&\cdots&\omega_{2d}^{2\cdot (d-1)}\\
\vdots&\vdots&\vdots&\ddots&\vdots&\vdots&\vdots&\ddots&\vdots\\
1&\omega_{2d}^{(d-2)\cdot2}&\cdots&\omega_{2d}^{(d-2)\cdot(d-2)}&\omega_{2d}^{(d-2)\cdot 1}&\omega_{2d}^{(d-2)\cdot 3}&\cdots&\omega_{2d}^{(d-2)\cdot (d-1)}\\
1&\omega_{2d}^{1\cdot2}&\cdots&\omega_{2d}^{1\cdot(d-2)}&\omega_{2d}^{1\cdot1}&\omega_{2d}^{1\cdot 3}&\cdots&\omega_{2d}^{1\cdot (d-1)}\\
1&\omega_{2d}^{3\cdot2}&\cdots&\omega_{2d}^{3\cdot(d-2)}&\omega_{2d}^{3\cdot1}&\omega_{2d}^{3\cdot3}&\cdots&\omega_{2d}^{3\cdot (d-1)}\\
\vdots&\vdots&\ddots&\vdots&\vdots&\vdots&\ddots&\vdots\\
1&\omega_{2d}^{(d-1)\cdot2}&\cdots&\omega_{2d}^{(d-1)\cdot(d-2)}&\omega_{2d}^{(d-1)\cdot1}&\omega_{2d}^{(d-1)\cdot3}&\cdots&\omega_{2d}^{(d-1)\cdot (d-1)}
\end{bmatrix}.
\end{align*}
\normalsize
Therefore $C^{(\ell_{even})}$ can be transformed by fundamental matrices as follows:
\begin{align*}
\overline C^{(\ell_{even})}=\begin{bmatrix}
1&1&1&\cdots&1\\
1&\omega_{2d}^{1\cdot1}&\omega_{2d}^{1\cdot2}&\cdots&\omega_{2d}^{1\cdot(d-1)}\\
1&\omega_{2d}^{2\cdot1}&\omega_{2d}^{2\cdot2}&\cdots&\omega_{2d}^{2\cdot(d-)}\\
\vdots&\vdots&\vdots&\ddots&\vdots\\
1&\omega_{2d}^{(d-1)\cdot1}&\omega_{2d}^{(d-1)\cdot2}&\cdots&\omega_{2d}^{(d-1)\cdot(d-1)}
\end{bmatrix}.
\end{align*}
Since $\overline C^{(\ell_{even})}$ is the Vandermonde matrix with size $d$, we get rank$(\overline C^{(\ell_{even})})=d$, and rank$(C^{(\ell_{even})})=d$. Hence by Lemma \ref{ncondi}, localization does not occur for the Fourier walk on the $d$-dimensional lattice for any $d\in\{1,2,3,\dots\}$. \hfill $\square$

\section{Summary}
We proved the non-existence of localization for the Fourier walk on the $d$-dimensional lattice $(d=1,2,3,\dots)$ by Theorem \ref{main}. To show this, we presented a necessary and sufficient condition and a necessary condition for the existence of localization for the space-homogeneous QWs by Lemma \ref{nscondi} and Lemma \ref{ncondi}, respectively. One of the interesting future problems might be to clarify the gap between the necessary and sufficient condition for the existence of localization and the necessary condition given by Lemma \ref{ncondi}.

\section*{Acknowledgments}
The author is grateful to Takeshi Kajiwara and Norio Konno for useful comments.


\begin{thebibliography}{9}
\bibitem{Aharonov1}
D. Aharonov, A. Ambainis, J. Kempe and U. V. Vazirani (2001), {\it Quantum walks on graphs}, Proceedings of ACM Symposium on Theory of Computation (STOC'01), July 2001, pp.50-59.

\bibitem{asanoetal}
M. Asano, T. Komatsu, N. Konno and A. Narimatsu, {\it The Fourier and Grover walks on the two-dimensional lattice and torus}, Yokohama Mathematical Journal  (in press). 

\bibitem{Kitagawa}
T. Kitagawa (2012), {\it Topological phenomena in quantum walks: elementary introduction to the physics of topological phases}, Quantum Inf. Process., {\bf11}, pp.1107-1148

\bibitem{komatsukonno}
T. Komatsu and N. Konno (2017), {\it Stationary amplitudes of quantum walks on the higher-dimensional integer lattice}, Quantum Inf. Process., {\bf 16}, 291

\bibitem{Komatsu and Tate.}
T. Komatsu and T. Tate (2019), {\it Eigenvalues of quantum walks of Grover and Fourier types}, J. Fourier Anal. Appl., {\bf 25}, pp.1293-1318

\bibitem{Konno1}
N. Konno (2008), {\it Quantum Walks}, Lecture Notes in Mathematics, {\bf1954}, pp.309-452, Springer

\bibitem{konnotakahashi}
N. Konno and S. Takahashi (2020), {\it On the support of the Grover walk on higher-dimensional lattices}, arXiv:2001.10261

\bibitem{Matsuoka}
L. Matsuoka, T. Kasajima, M. Hashimoto and K. Yokoyama (2011), {\it Numerical study on quantum walks implemented on the cascade rotational transitions in a diatomic molecule}, J. Korean Phys. Soc., {\bf 59}, pp.2897-2900

\bibitem{Portugal}
R. Portugal (2018), {\it Quantum Walks and Search Algorithms}, second edition, Springer 

\bibitem{stefanak}
M. Stefanak, B. Kollar, T. Kiss and I. Jex (2010), {\it Full revivals in 2D quantum walks, Physical Scripta}, {\bf 2010}, T140

\bibitem{Tate}
T. Tate (2014), {\it Eigenvalues, absolute continuity and localizations for periodic unitary transition operators}, arXiv:1411.4215 

\bibitem{watabeetal}
K. Watabe, N. Kobayashi, M. Katori and N. Konno (2008), {\it Limit distributions of two-dimensional quantum walks}, Phys. Rev. A, {\bf 77}, 062331

\end{thebibliography}
\end{document}